# A Multi-layered GaGeTe Electro-Optic Device Integrated in Silicon Photonics

Srinivasa Reddy Tamalampudi[1,*], Ghada Dushaq[1], Juan Esteban Villegas[1,2], Bruna Paredes[1], Mahmoud S. Rasras[1,*]

[1]Department of Electrical Engineering, New York University Abu Dhabi, UAE

[2]Department of Electrical and Computer Engineering, New York University Tandon School of Engineering, NY, 11201, Brooklyn, USA

* st4212@nyu.edu, mr5098@nyu.edu

*Abstract*— Electrically tunable devices contribute significantly to key functions of photonics integrated circuits. Here, we demonstrate the tuning of the optical index of refraction based on hybrid integration of multi-layered anisotropic GaGeTe on a silicon micro-ring resonator (Si-MRR). Under static applied (DC) bias and transverse-electric (TE) polarization, the device exhibits a linear resonance shift without any amplitude modulation. However, for the transverse-magnetic (TM) polarization, both amplitude and phase modulation are observed. The corresponding wavelength shift and half-wave voltage length product ($V_\pi.l$) for the TE polarization are 1.78 pm/V and 0.9 V.cm, respectively. These values are enhanced for the TM polarizations and correspond to 6.65 pm/V and 0.28 V.cm, respectively. The dynamic radio frequency (RF) response of the devices was also tested at different bias conditions. Remarkably, the device exhibits a 1.6 MHz and 2.1 MHz response at 0 V and 7 V bias, respectively. Based on these findings, the integration of 2D GaGeTe on the silicon photonics platform has great potential for the next generation of integrated photonic applications such as switches and phase shifters.

## I. INTRODUCTION

Opticl phase shifters (PS) and modulators are essential components in modern communication, microwave photonics, and quantum photonic systems [1]. $LiNbO_3$ crystals constitute most standard external modulators [2]. However, to satisfy the future data communication bandwidth requirements, compact on-chip integration of the modulators is highly desired [3, 4]. Despite the recent attempts to integrate $LiNbO_3$ waveguide modulators, their dimensions are still in millimeters scale. Additionally, they still exhibit high half-wave voltage product ($V_\pi.l$) which increases the overall modulation power consumption [5, 6].

On the other hand, Si-based optical modulators suffer from low phase-modulation efficiency owing to the weak plasma dispersion effect in Si, which results in large optical losses [7]. In comparison, hybrid integrated III–V materials on silicon provide a good balance of modulation bandwidth and efficiency. They rely on the plasma dispersion effect where the lighter n-doped/ heavily p-doped (> $10^{19} cm^{-3}$) III–V materials result in a stronger electron/hole-induced refractive index change when compared to silicon [8]. However, since the plasma-dispersion effect can only induce small variations in the index of refraction, interaction lengths in the range of hundreds of microns to millimeters are required [9]. While many materials have been explored as electro-optic devices on a silicon platform, atomically thin two-dimensional (2D) materials offer a strong foundation for broadband on-chip electro-optic devices operating from ultraviolet to far-wave infrared range [10].

In particular, the birefringence in anisotropic 2D materials allows controlling the real part (Δn) and imaginary part (Δk) of the refractive index by controlling the crystal orientation with respect to the polarization state of the propagating light [11]. Recently, black phosphorus (BP) waveguide integrated electro-optic modulators have been demonstrated which exhibited a polarization-dependent absorption and extinction ratio (ER)[12]. Furthermore, I. Datta *et.al.* demonstrated a capacitive $WSe_2/Al_2O_3$/graphene-type modulator [13]. In this structure, the $WSe_2$ and graphene served as electro-refractive and electro-absorptive materials, respectively to tune the Δn/ Δk ratio (~1) and reduce the device insertion losses. However, power consumption is an important figure of merit in electro-optic devices, hence, identifying new materials that allow controlling of the Δn/ Δk ratio, small footprint, and low $V_\pi.l$ are highly desired [8].

Recently, 2D-GaGeTe crystals have gained a lot of attention due to their superior electrical and optical properties, in addition to their high chemical stability under ambient conditions [14-17]. The GaGeTe crystals are part of the MXTe (M = Al, Ga; X = Si, Ge, Sn) 2D layered materials group which are made up of stacks of monolayers connected by weak van der Waals interactions [18]. Remarkably, recent studies on GaGeTe crystals identified the existence of the non-centrosymmetric structure[19]. Here, we present an electro-optic device based on the hybrid integration of GaGeTe into a silicon micro-ring resonator (Si-MRR). For the TE polarization, pure phase modulation is observed which is mainly attributed to the change in the real part of the index of refraction. However, a combination of phase and amplitude modulation is observed for devices operating in the TM Mode. The GaGeTe/Si-MRR exhibits a blue-shifted phase with positive applied bias, low $V_\pi \cdot l$, and small footprint.

The paper is structured as follows, Section II discusses the design of the electro-optic device, fabrication, and the optical properties of the GaGeTe film. The interaction of the light with the 2D film, the effect of polarization, and voltage dependence are discussed in Section III. Section IV presents a discussion of the underlying effects of phase shifting and absorption. The RF response is evaluated in Section V. Finally, conclusions are given in Section VI.

## II. DESIGN AND FABRICATION

### A. The Optical Parameters of GaGeTe

To validate the anisotropic nature of the GaGeTe flakes (70 nm thick), Angle-Resolved Polarized Raman Spectroscopy (ARPRS) measurements are carried out. The spectra are recorded for two configurations: parallel polarization (the incident polarization angle (P) is the same as the analyzer's angle (A)) and perpendicular polarization (P and A are perpendicular). Figure. 1a shows the variance of the Raman peak intensities with the incident light polarization which confirms the in-plane optical anisotropy of the GaGeTe flakes [20]. Additionally, the measured Raman peak positions are in agreement with other reports [19].

Next, we measured the real ($n$) and complex ($k$) optical parameters of the GaGeTe flakes for the in-plane (xx) and out-of-plane (zz) orientations using high-resolution ellipsometry. Results are shown in Fig. 1b. Best to our knowledge, this is the first experimental demonstration of the GaGeTe crystal optical parameters. The observed dependence on the plane orientation is due to the crystal axes-dependent optical refractive index which further confirms the anisotropic properties of the GaGeTe. As can be seen, for visible (VS) to short-wave infrared (SWIR) (450 nm to 1660 nm) wavelengths, the real part $n$ is higher in the in-plane compared to the out-of-the-plane direction. In contrast, in this wavelength range, the imaginary part $k$ is lower for the in-plane direction compared to the out-plane orientation. At 1310 nm wavelength, the extracted in-plane values are $n$ =3.68 and $k$= 0.25 which are key values needed to determine the light confinement in the GaGeTe layer of the waveguide/2D phase shifter structure.

### B. Device Design & Fabrication

The devices are fabricated using SOI wafers with a 220 nm top silicon layer and a 2 μm buried oxide layer. To evaluate the index of refraction tuning performance, we used an integrated micro-ring resonator (MRR) Si/GaGeTe structure, (see Fig. 2). The waveguide width of the MRR cavity is 460 nm with a ring radius of 40 μm. A bus waveguide couples light to the MRR through a 100 nm gap spacing then collected at the output via through port. The optical coupling between the fiber and the chip was achieved using a lensed fiber. To test the device's static and dynamic performance, three metal pads of Ti/Au (10 nm/120 nm) were deposited symmetrically as electrodes. Notably, the deposited electrodes and the Si waveguide are not in the same plane and there is a 100 nm height difference between them, (see Fig. 2a).

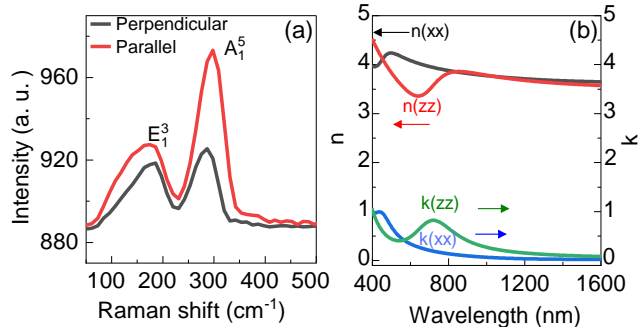

Fig. 1: (a) Angle Resolved Polarized Raman characteristics of the GaGeTe flake. (b) the optical parameters of the GaGeTe in the in-plane (xx-ordinary) and out-of-plane (zz-extraordinary) crystal directions.

A deterministic dry transfer procedure was used to transfer the multilayer GaGeTe on top of the Si-MRR and pre-deposited metal electrodes [21]. Figure 2b shows an optical microscope image of the fabricated MRR where the exfoliated GaGeTe flake is highlighted with a red dashed box. A scanning electron microscopy (SEM) image of the device is shown in Fig. 2c. As can be seen, the flake is perfectly aligned and adhere conformably to the photonic structure beneath it. Moreover, since the GaGeTe flake has a higher refractive index ($n$ ~ 3.7) than Si ($n$ ~ 3.4) at 1310 nm, a large portion of the optical mode is confined into the GaGeTe multilayer. The metal electrode's gap is fixed at 8 μm, which minimizes the optical absorption losses. The simulated guided optical modes for the quasi-TE/TM polarizations with and without flakes are shown in Fig. 2d. The FDTD simulations are performed for a GaGeTe layer thickness of 70 nm at 1310 nm wavelength. It is important to note that for both polarizations, the guided optical mode of the Si waveguide overlaps with the GaGeTe flake.

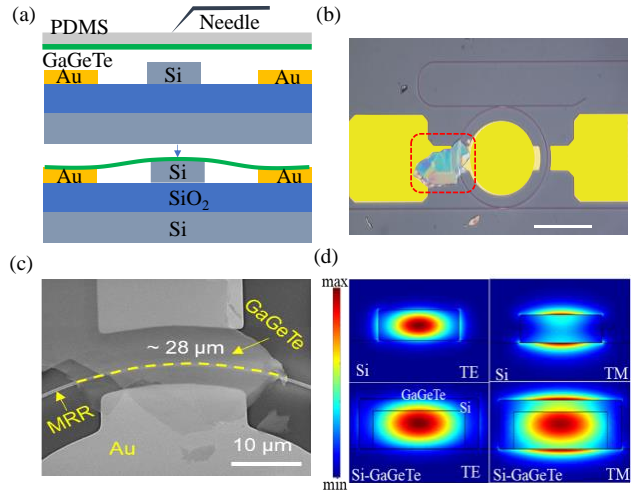

Fig. 2. (a) Schematic representation of the deterministic transfer of the GaGeTe flake onto the Si-MRR. (b-c) Optical microscopy and SEM images of the Si-MRR/GaGeTe device. (d) Guided light-multilayer GaGeTe interaction: Electric-field profiles ($|E|^2$) of TE and TM modes of bare Si waveguide (top panel) and 70 nm GaGeTe on Si (bottom panel) at 1310 nm.

III. OPTICAL TRANSMISSION MEASUREMENTS

To verify the phase shift and index of refraction tuning of the hybrid waveguide structure, we measured the optical transmission spectra of the device in the wavelength region of 1300 nm-1320 nm for both polarization modes. The design is fabricated to target a high Q-factor value ($\lambda$/FWHM~11750). In the experiment, a lensed fiber is used to side couple light from a tunable laser (Keysigth 81606A) into and off the SiPh chip. The input light polarization is set using external fiber-based waveplate paddles. Light is then detected using an optical power meter (Keysight-N7744A).

Figure 3a depicts the TM transmission spectra before and after integrating the 70 nm flake that has a 28 µm interaction length. The Si-MRR initial resonance dip loss (RDL) is ~20 dB and it has a free spectral range (FSR) of 1.045 nm. After integrating the GaGeTe flake, The MRR's round-trip loss has increased due to light absorption and edge scattering. Additionally, the RDL of Si-MRR/GaGeTe is reduced to ~10 dB while the measured FSR is increased to 1.33 nm with a resonance shift to a longer wavelength. Similarly, the transmission spectra for the TE-polarized light are shown in Fig. 3b. The RDL change remains within $\pm 0.2$ dB (limited by measurements resolution) and the FSR is reduced from 1.58 nm to 1.55 nm. Note that for the TM case, losses due to absorption in the flake reduce the Q-factor of the resonator.

*A. Electro-Optic Tuning*

To determine the influence of the applied bias on the resonance wavelength shift, the device response is investigated in the optical O-band. To eliminate any impact of the input laser power, first the lunched power to the Si chip is varied from 0 to 7.5 mW while the applied bias to the device is kept at 0V. We observed no shift in the resonance peak positions for both polarizations, (see Figs. 3(c-d)). Hence, the propagating light induces no thermal dissipation in the device. Thus, in subsequent measurements, the laser power is kept below 7.5 mW.

The transmission spectra for the TM polarization under different bias voltages are shown in Fig. 4(a). A blue shift in the resonance wavelengths occurs as the applied bias increases. A linear shift of 66.5pm is observed up to 10 V bias and it increases nonlinearly to 224.8 pm at 14 V bias. Since thermal heating results in a red shift in the Si-MMR resonance wavelength [22], the observed blue shift indicates an electro-optic modulation of the refractive index of the combined Si/GaGeTe guiding structure[23]. Note that the dissipated electrical power due to the dark current and applied voltage are 1.02 mW and 1.99 mW at 10V and 14 V, respectively.

Figure 4b depicts the resonance shift versus voltage extracted from the linear region of the phase shift. A high tuning efficiency of $(\alpha) = \Delta\lambda/\Delta V \approx$ -6.65 pm/V is measured. However, despite the linear phase shift, small amplitude modulation of the resonance dips is also noticed which indicates a simultaneous change in $\Delta k$. As the bias increases to 14 V, a significant shift in resonance is observed combined with a larger adjustment in the resonance dip.

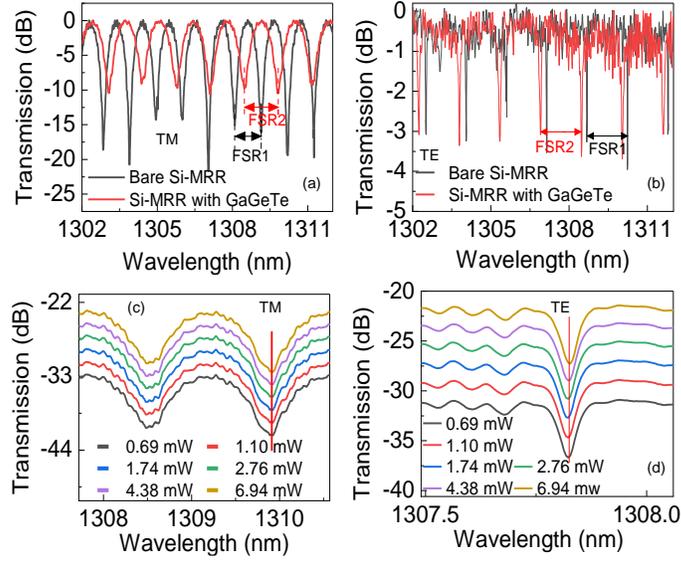

Fig. 3. (a-b) The normalized transmission spectra of the MRR without (black) and with (red) the GaGeTe flake for the TM and TE modes, respectively. (c-d) Transmission spectrum of the Si-MRR with GaGeTe under various propagating light intensities for (c)TM and (d) TE polarizations, respectively.

Figure 4c depicts the resonance peak shift at different applied bias voltages (0 to 11V) for TE polarization. A pure blue phase shift is observed with no change in the resonance peaks (dips). The measured resonance peak shift is 26.1 pm at 11 V. As depicted in Fig. 4d, a linear shift is observed similar to that of LiNbO$_3$ phase modulators. The measured tuning efficiency ($\alpha$) = $\approx$ -1.78 pm/V.

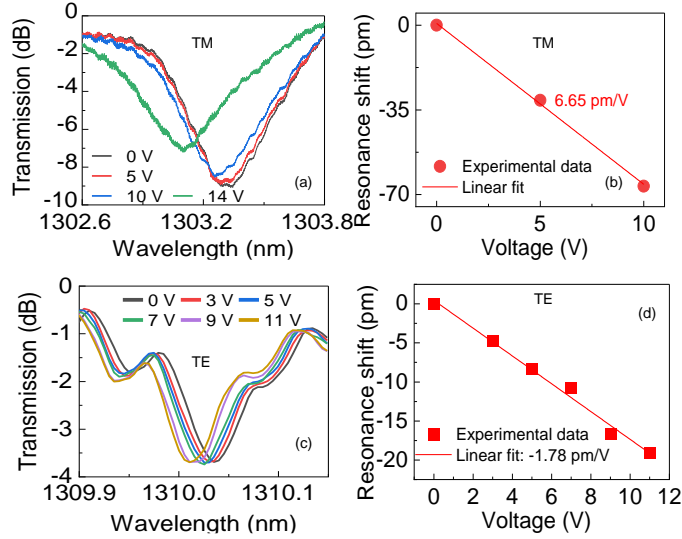

Fig. 4. (a) The electro-static spectral response of the hybrid Si-MRR/GaGeTe flake for TM mode with coverage lengths of 28 μm for different DC voltages. (b) Measured resonance shift versus applied voltage and the corresponding linear fitting curves of all three devices for TM mode. (c) The electro-static spectral response of the hybrid Si-MRR/GaGeTe flake for different DC voltages measured for TE polarization (d) measured resonance shift versus applied voltage and the corresponding linear fitting curve of the device operating in TE mode.

## B. Phase-Shift Figure of Merit

The tuning figure of merit (FOM) of the Si-MRR integrated phase modulators can be expressed as a $V_\pi \cdot l$ product of $|l_{GaGeTe} \cdot \lambda_{FSR} \cdot dV/2d\lambda|$. [24, 25] The length of the active waveguide region is $l = 28$ μm while $V_\pi$ is the voltage required to induce a π shift phase in the waveguide. The measured $V_\pi \cdot l$ values for the TM (TE) polarization for a FSR = 1.33 nm (1.58 nm) are 0.28 V·cm (1.24 V·cm). The measured results are compared with traditional phase modulators and summarized in Table 1.

TABLE 1.
HYBRID INTEGRATED ELECTRO-OPTIC PHASE MODULATORS

| Material/platform | Type | Size (μm) | Efficiency (pm/V) | $V_\pi \cdot l$ (V·cm) | Ref. |
|---|---|---|---|---|---|
| LiNbO$_3$/SOI | MZ | 5000 | - | 2.55 | [26] |
| BaTiO$_3$/Si$_3$N$_4$ | MZ | 1000 | - | 0.33 | [27] |
| PZT/ Si$_3$N$_4$ | MRR | 524 | -9.98 | 3.2 | [24] |
| GaGeTe/SOI | MRR | 28 | -6.65 (TM) | 0.28 | This work |
|  |  |  | -1.78 (TE) | 1.24 |  |

## C. Insertion Loss and Amplitude Modulation

To explore the use of the proposed phase shifter as an optical modulator, we analyze the Si-MRR insertion loss during the phase tuning process. As shown in Fig. 5a, the resonance losses for the TM mode decrease from 8.45 dB (at 5 V) to 6.4 dB (at 14 V) at a probe wavelength of $\lambda_0 = 1303.3$ nm. At this maximum bias, the cavity also experiences a phase change of $\sim \frac{\Delta\lambda}{FSR}\pi = \pi/3$ (= 224.8/ 1330.0 π) which corresponds to a simultaneous change in Δn and Δk.

A change in the imaginary part of the complex refractive index alters the resonance loss because it is linked to the optical absorption, while a change in the real part manifests itself as a shift in the transmission spectrum. In our measurements, for a bias voltage from 0 to 10 V, the device operates in the under-coupling condition. This remains true as the applied voltage mainly alters the real part of the index of refraction where a tiny variation in the resonance losses is due to a small absorption. However, at higher voltage (14 V), the increase in free carriers results in a significant increase in the imaginary part along with the real component of the index. Similar behavior is observed in silicon EO modulators [28].

Figure 5b illustrates the modulation depth of the device when operating at selected wavelengths from 1302.6 nm to 1303.8 nm. Their spectra are compared under a bias of 10 V and 14V. A modulation depth larger than 3dB is obtained over a wavelength range of 0.45 nm, $\Delta\lambda_1$ (1302.91 nm to 1303.05) and $\Delta\lambda_2$ (1303.26 nm to 1303.25 nm), in which the largest modulation depth can reach up to 4.6 dB. The "OFF" state of the ring modulator is defined as the lowest transmission at the resonant wavelength while the maximum transmission defines the "ON" state. It is observed that modulation of 47.5% (3.47 dB) can be achieved with a bias difference of 4 V (from 10 V to 14 V). In addition to the blue shift in the resonance wavelength, at this bias, the resonance condition is switched from under-coupled to over-coupled.

On the other hand, for TE polarization, the device exhibits a small reduction of 0.25 dB insertion losses (see Fig 4c) when the applied DC voltage changes from 3V to 11 V. The voltage-controlled insertion loss for the TE mode is negligible.

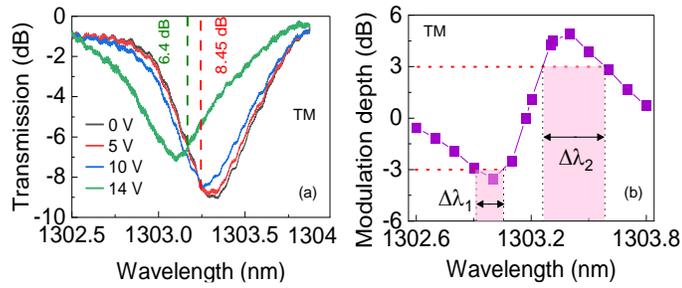

Fig. 5. (a) Normalized transmission spectrum of the Si-MRR with GaGeTe device under different bias voltages from 0 and 14 V respectively. (b) The modulation depth at various wavelengths is measured by comparing transmission spectra at 0 V and 14 V. The vertical pink color dashed area indicates the wavelength range of 0.45 nm over which a modulation depth greater than 3 dB was achieved.

## IV. THE UNDERLAYING EO EFFECTS: DISCUSSION

The phase shift and absorption can be explained by studying the propagation of the optical mode in the hybrid Si/GaGeTe waveguides. The polarization-dependent guiding condition for the optical modes can be given by the following equations [29],

For TE:

$$1 - \frac{2\pi i \sigma(\omega)}{\omega\sqrt{q^2 - \omega^2/c^2}} = 0 \qquad (1)$$

For TM:

$$1 + \frac{2\pi i \sigma(\omega)\sqrt{q^2 - \omega^2/c^2}}{\omega} = 0 \qquad (2)$$

where c is the speed of light and σ(ω) is the local dynamic conductivity of the 2D gas, $i = \sqrt{-1}$ is the complex number, ω is the light angular frequency and q is the electric charge. Equations (1) and (2) lead to the conclusion that the imaginary part of the conductivity governs the light absorption whereas the stronger absorption for the TM mode indicates a negative conductivity imaginary part for the GaGeTe film. Additionally, the alignment of the anisotropic GaGeTe crystal axis (zig-zag/ armchair or in between) with respect to the propagating light polarization plays an essential role in the absorption of the propagating light [12].

Moreover, when a material is subjected to an external electric field, an electro-optic (EO) effect induces changes in the real ($n$) and imaginary ($k$) parts of its index of refraction[30]. A linear change to the real component is referred to as the Pockels effect (proportional to the electric field $E$) while the Kerr effect is proportional to $E^2$ describes second-order nonlinearity. Other effects include the plasma dispersion (PD) effect which is related to the density of free carriers in a semiconductor and the Franz–Keldysh (F-K) effect which results in a change in the optical absorption in a semiconductor when an external electric field is applied. [30]. The latter two effects change both the real and imaginary parts of the refractive index.

In our study, the devices exhibit a linear blue phase shift for the TE polarization without experiencing any intensity modulation. Therefore, this rules out the non-linear Kerr effect. Additionally, since no induced losses are observed, the PD and F-K effects are also excluded. Therefore, it is evident that the Pockels effect dominates these devices. As for the TM polarization, the devices showed a simultaneous change in the $n$ and $k$ values. The phase shift is similar to the TE mode but has a larger coefficient shift. Additionally, a small loss is observed for small voltages. However, a significantly larger blue shift and higher attenuation are observed at larger applied voltages. This indicates a combination of the Pockels and plasma dispersion effects in the case of TM polarization.

Furthermore, the recorded polarized Raman spectroscopy and ellipsometry results provide evidence of the GaGeTe crystal's anisotropy, where unique $n$ and $k$ values in the in-plane and out-of-plane crystal orientations are observed. Hence, if the change in the crystal's anisotropic permeability tensor's components is linear with the externally applied electric field, the EO effect is the Pockels effect. It is worth noting that only non-centrosymmetric crystals exhibit the Pockels phenomenon. Recent studies on grown GaGeTe flakes demonstrated a non-centrosymmetric structure [19]. Another point to consider is the height difference between the Si waveguide and metal electrodes which induces a strain effect in the transferred GaGeTe. It was well reported that strain can induce the Pockels effect in centrosymmetric silicon crystal[31].

## V. Dynamic Electro-Optical Modulation

To measure the radio frequency (RF) response of the devices, a continuous-wave (CW) 1310 nm laser is coupled to the Si-MRR device modulated by applying an RF signal generated from a lock-in amplifier (UHFLI 600). A static bias and RF signal are applied to the GaGeTe devices using a Bias-T. An external photodiode is used to detect the modulated light signal, which is then pre-amplified and fed back to the lock-in amplifier.

Figure 5 depicts the frequency response of the phase shifter when biased at 0 V and 7 V. The measured 3dB bandwidths are 1.6 and 2.1 MHz, respectively (RF $V_{pp}$ = 0.75 mV). Note that this observed electro-optic modulations of a few MHz hents towards the Pockels effect which exhibits a shorter timescale.

The total response time of the devices is determined by the carrier transient time ($\tau_{tr}$), and the junction capacitance charge/discharge time constant ($\tau_{rc}$). The former can be expressed as $\tau_{tr} = l^2/2\mu V_{ds}$, where $l$, μ, $V_{ds}$ are the length of the GaGeTe channel, mobility of a GaGeTe flake, and bias voltage. The measured 3 dB bandwidths vary with the applied bias voltage (See Fig. 5), suggesting that $\tau_{tr}$ determines the total response time ($\tau$). As a result, reducing the channel length and shortening the transient time help to achieve a larger bandwidth. In future designs, impedance matching of the GaGeTe devices is necessary to obtain a larger bandwidth.

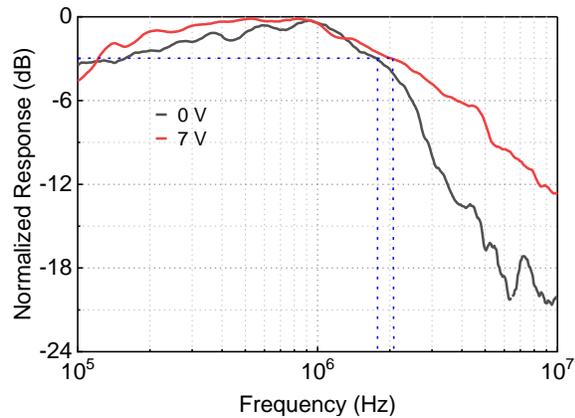

Fig. 5. (a) Radiofrequency response with a 3-dB bandwidth of 1.6 MHz at zero bias voltage and 2.1 MHz at 7 V.

## VI. Conclusion

The study analyzes the hybrid integration of GaGeTe onto a silicon photonics platform to construct an electro-optic phase tuning element. The constructed device is simple with two-terminal which eases the fabrication process. Wide phase shift tuning is achieved using a short interaction length of 28 µm. This footprint is smaller than conventional integrated $LiNbO_3$ modulators (typically a few mm). Furthermore, the measured half-wave voltage length $V_\pi \cdot l$ product (see Table 1) outperforms that of $LiNbO_3$ or $BaTiO_3$ modulator. Based on our results, it is evident that the Pockels effect is the dominant phase tuning factor for the TE polarized light, while a combination of the Pockels and plasma dispersion effects existed in devices for the TM polarization.

To conclude, the hybrid integration of GaGeTe on Si waveguides expands the silicon photonics device library by introducing a new class of highly tunable and linear electro-optical materials. The proposed structure can be used to construct optical phase shifters, high-speed modulators, small-footprint switches, and a variety of other active integrated photonic devices.

## Acknowledgment

This work was supported by NYUAD Research Enhancement Fund. The experimental characterization was conducted in the NYUAD Photonics Lab and simulations were performed on the NYU IT High-Performance Computing resources given its services and staff expertise. We are very thankful to Nikolas Giakoumidis and the Core Technology Platform Facility (CTP) for all the technical and instrumentation support. We also acknowledge the support of Dr. James Weston and Dr. Qiang Zhang for instrumentation support. S. R. T and G. D contributed equally to this work. G.D acknowledges the support from L'Oréal UNESCOs for women in science.